\documentclass[a4paper,fleqn,usenatbib,referee,epsfig,color]{mn2e}

\usepackage[T1]{fontenc}
\usepackage{ae,aecompl}
\setcitestyle{numbers,open={[},close={]}}
\usepackage{graphicx}	
\usepackage{amsmath}	
\usepackage{amssymb}	
\usepackage{subfig}
\usepackage{bm}
\newif\ifAMStwofonts
\newcommand{\be}{\begin{equation}}
\newcommand{\ee}{\end{equation}}
\newcommand{\bea}{\begin{eqnarray}}
\newcommand{\eea}{\end{eqnarray}}
\newcommand{\nn}{\nonumber}

\newcommand{\ket}{\mid\bm{\Psi~}\rangle}

\title[Dirac/Clifford Relativity]{\bf  Quantum Equations of Motion and the Geometrical Imperative : Relativistic}
\author[R.N. Henriksen ]
{R. N. Henriksen$^1$\thanks{henrikr@queensu.ca} \\ 
$^1$Dept. of Physics, Engineering Physics \& Astronomy, Queen's University, Kingston, Ontario, K7L 3N6, Canada\\}

\pubyear{2021}
\begin{document}
\label{firstpage}
\pagerange{\pageref{firstpage}--\pageref{lastpage}}
\maketitle
\begin{abstract}
We extract the square root of the Minkowski metric using Dirac/Clifford matrices.  
The resulting  $4\times 4$ operator $d{\bf S}$ that represents the square root, can be used to transform four vectors between relatively moving observers. This effects the usual Lorentz transformation. In addition it acts on a Dirac bi-spinor. The operator is essentially a Hamiltonian that can be used to write an equation of motion for a relativistic spinor. This turns out to be the Dirac equation for electrons in standard form.
We believe that is is a new approach to familiar results.
\end{abstract}

\begin{keywords}
Special Relativity, Dirac Equation, Spinors
\end{keywords}

\section{Introduction}
The squared distance interval in Minkowski space-time is not positive definite.  A general extraction of the square root of the spacetime interval regardless of sign does not seem to have been studied for possible physical insight. However some authors \cite{K2008}, \cite{BK2008} have used an augmented Clifford algebra based on Pauli matrices to describe spacetime based on what they call the "Algebra of Real Space'"(ARS). This leads them to a similar deduction of the Lorentz transformations and of the Dirac equation. However the group theory approach is neither very explicit nor intuitively direct.

Effectively, the square root of the metric of spacetime defines  a  matrix  that can operate either on four vectors or on bi-spinors. It is equivalent to the Feynman `slash" when operating on four vectors.
 We find a matrix representation of the square root  by using a Dirac/Clifford Algebra. We study the action both on four vectors and on bi-spinors below. The square root extraction is readily extended to  curved space-times with diagonal metric tensors.  This becomes a convenient way to formulate the motion of quantum particles near a (non rotating) black hole, although for spinors the equations can be formidable.


We find that the action of the square root matrix  provides a straight forward derivation of the Lorentz transformation of four vectors, but also of the Dirac equation, when quantized and  allowed to act on a four component  wave function. This process is strictly parallel to the deduction of the Pauli equation that follows by starting from the Euclidian metric.

The argument shows that both classical and quantum equations of motion of a free particle can be found by transforming to the rest frame of the particle by using this matrix. This resembles the r\^ole of the Hamiltonian operator in producing a `flow', but the motion is more explicit.  We believe that this description offers insight into the physical relation between  spacetime and the equation of motion of a  quantum particle.  

In section (\ref{sect:extract}) we carry out the square root of the metric operation, and define the  matrix that is the essence of this study. In section (\ref{sect:properties}) we modify the matrix slightly to form the `velocity matrix' and study its properties. 

In section (\ref{sect:Dirac}) we come at last to the Dirac equation. We show that by quantizing the velocity matrix it becomes  the Hamiltonian operator, and allowing it to operate on a Dirac four spinor of a free particle gives the quantum time dependence of the spinor. This relation is equivalent to the Dirac equation.  The matrix functions as a Hamiltonian operator, but it also suggests that the Dirac equation results from a quantized Lorentz transformation to the rest frame of the spinor. We indicate the changes necessary to place the bi-spinor in a constant external magnetic field and give an explicit equation.

\section{Extracting the Square Root}
\label{sect:extract}
We use  signature such that the squared interval takes the form ($c=1$ for the moment) 
\be
ds^2=dt^2-d{\bf r}^2,\label{eq:interval}
\ee
where $d{\bf r}$ is the spatial interval. We will normally use Cartesian spatial coordinates, so that the spatial interval is the positive definite value $d{\bf r}^2=dx^2+dy^2+dz^2$. Curvilinear coordinates, including those imposed by General Relativity (GR), require slightly different Dirac/Clifford matrices. 

Although our approach features the metric of space-time, it is useful to remember that when following the world line of a moving point $(ds/dt)^2=1/\gamma^2$.
This is related to the energy $E=\gamma m$ and the momentum ${\bf p}=\gamma m{\bf v}$  of a moving point mass through 
\be
\gamma^2\equiv \frac{1}{1-v^2}=\frac{E^2}{E^2-p^2}.
\ee
Thus the link between the space-time metric and relativistic dynamics of a massive particle is rather direct.

In fact, in an almost unique approach \cite{Harris1972}, the relativistic expression for the energy squared-momentum squared relation has its square root extracted using  Dirac/Clifford matrices. When  combined with the quantum energy and momentum operators the Dirac equation follows. What is new in this paper, is the emphasis on connecting the square root of the metric to both the Lorentz transformation and the Dirac equation. 

To extract the square root of the metric we assume the square root in the form
\be
ds{\bf I}=\bm{\gamma}_odt-\bm {\gamma}_jdx^j\equiv \bm{\gamma}_adx^a\equiv d{\bf S},\label{eq:squareroot}
\ee
where the $\bm {\gamma}_\alpha$ are Dirac/Clifford matrices. We use the summation convention, and Latin letters before `h' are summed $0..3$ and those after `h' are summed $1..3$. 

In order for this expression to represent the square root of the metric, the Dirac/Clifford matrices must satisfy 
\bea
\bm{\gamma}_0^2&=&{\bf I},\nonumber\\ 
\bm{\gamma}_0\bm{\gamma}_i&+&\bm{\gamma}_i\bm{\gamma}_0=0,\nn\\
\bm {\gamma}_i\bm{\gamma}_j&+&\bm{ \gamma}_j\bm{\gamma}_i=-2\bm{I}\delta_{ij},\label{eq:gammamatrices},
\eea
where $\bm{I}$ is the four dimensional unit matrix. A direct calculation of the square of expression (\ref{eq:squareroot}) yields 
\be
ds^2{\bf I}=(dt^2-d{\bf r}^2){\bf I},\label{eq:expsquareroot}
\ee
so that the square root has been taken by the summed matrix on the right of expression (\ref{eq:squareroot}).

We choose the Dirac/Clifford matrices to be (in $2\times 2$ block form)

\bea
\bm{\gamma}_0=\left(\begin{array}{clcr}&{\bf 1} & {\bf 0} \\&{\bf 0} & -{\bf 1} \end{array}\right), ~~ \bm{\gamma}_1=i\left(\begin{array}{clcr}&{\bf 0}  &\bm{\sigma}_1\\&\bm{\sigma}_1 &  {\bf 0}\end{array}\right)\label{array:gammas}
 \eea

and 
\bea
\bm{\gamma}_2=i\left(\begin{array}{clcr}&{\bf 0} & \bm{\sigma}_2 \\&\bm{\sigma}_2 & {\bf 0} \end{array}\right), ~~ \bm{\gamma}_3=i\left(\begin{array}{clcr}&{\bf 0}  &\bm{\sigma}_3\\&\bm{\sigma}_3 &  {\bf 0}\end{array}\right)\label{array:2ndgammas}
\eea
In these expressions ${\bf 1}$ and $\bf{0}$ are the $2\times 2$ unit  and zero matrices respectively. 
\vspace{1cm}

The $\bm{\sigma}_j$ are the Pauli spin matrices namely
\bea
\bm{\sigma}_1=\left(\begin{array}{clcr}&0 &  1 \\& 1 & 0\end{array}\right),~~\bm{\sigma}_2=\left(\begin{array}{clcr} &0 & -i \\& i & 0 \end{array}\right),~~\bm{\sigma}_3=\left(\begin{array}{clcr} &1 & 0\\& 0 & -1\end{array}\right)\label{array:Pauli1}
\eea

In a standard reference such as \cite{AH1982}, our notation corresponds to their $\bm{\alpha}_j$ and $\bm{\beta}$ matrices according to
\be
\bm{\gamma}_0=\bm{\beta},~~~~~~~~~~\bm{\gamma}_j=i\bm{\alpha}_j.\label{eq:standardref}
\ee

The matrices (\ref{array:gammas}), (\ref{array:2ndgammas}) are slightly non-standard when it comes to quantization, due to the explicit presence of $-i$ as a factor in the spatial part of equation (\ref{eq:squareroot}). 

With these definitions equation  (\ref{eq:squareroot}) implies that  for Minkowski space-time 
\be
~~~~~~~~~~~~~~~~~~~~~~~~~~~~~~ds{\bf I}=d{\bf S}\equiv \bm{\gamma}_adx^a\equiv \bm{\beta}dx^0-i\bm{\alpha}_jdx^j,\label{eq:correspondence}
\ee
where the summed matrix is
\bea
d{\bf S}=\left(\begin{array}{clcr}dt&0&-idz&-D_+\\0&dt&D_-&idz\\-idz&-D_+&-dt&0\\D_-&idz&0&-dt\end{array}\right)\label{array:dS}
\eea
In equation (\ref{eq:correspondence}) we have also given the matrix in terms of the $\bm{\beta}$ and $\bm{\alpha}_j$ matrices of \cite{AH1982}. The matrix is the differential form (the differentials are numbers, not 1-forms) that denotes the general square root of the Minkowski metric.
We refer to it briefly as the {\it root} matrix. 
The differential operators  in expression (\ref{array:dS}) are defined as
\be
D_{\pm}=dy\pm idx.\label{eq:Dfoms}
\ee

It is readily verified that (standard matrix multiplication is implied for adjacent matrices)

\be ~~~~~~~~~~~~~~~d{\bf S}^2\equiv d{\bf S}d{\bf S}=ds^2{\bf I}.\label{array:dS2}
\ee
This property is invariant under a cyclic permutation of $\{xyz\}$, so that the root operator and its derivative can take on three different forms. These correspond to three orientations of the spatial axes while maintaining the right handed sense.

This $4\times 4$ matrix operator can operate either on a four vector in Minkowski space or on a bi-spinor in Hilbert space (once quantized). 
Acting on a four vector, ${\bf A}$, the matrix $d{\bf S}$ gives another four vector as 
\be
~~~~~~~~~~~~~~~~~~~~~~~~~~~~~~d{\bf B}=d{\bf S}{\bf A}.\label{eq:trans1}
\ee


Hence $d{\bf S}^2$, operating on any four vector ${\bf A}$ in Minkowski space, gives $ds^2$ times the vector ${\bf A}$.   

The question that arises is, what can the significance  of the operation by $d{\bf S}$ be? That is, what is $d{\bf B}$ given ${\bf A}$? We explore this at length  in the sections below, first classically and then quantum mechanically. Clearly we will want to study the transformation in a dimensionally coherent form.

 It should be noted here that $d{\bf S}$ can also be used in integral form, starting the integration from the space-time origin. This  turns the differentials to their integrated coordinate form.  Minkowski spacetime is a real space, but we will treat it as a complex space by writing  $i$ in front of spatial coordinates  of a four vector. Therefore, the magnitude of a four vector is found by summing the squares of the components.

\subsection{Properties of the Transformation Matrix}
\label{sect:properties}

We will work mainly with the derivative of the root matrix ${\bf S}$ in the form 
\be
~~~~~~~~~~~~~~~~~~~\dot {{\bf S}}\equiv \frac{d{\bf S}}{ds},\label{array:Sdot}
\ee
which we refer to briefly as the {\it transformation} matrix. Between inertial observers $ds$ is the invariant proper length, so that this matrix is dimensionless. 

Applying this matrix  we will find a new four vector ${\bf B}$ given ${\bf A}$, according to  
\be
~~~~~~~~~~~~~~~~~~~~{\bf B}=\dot{{\bf S}}{\bf A}.\label{eq:trans2}
\ee

The expression (\ref{array:dS2}) now shows that the transformation matrix is its own inverse, that is 
\be
~~~~~~~~~~~~~~~~~~~~\dot{{\bf S}}={\dot{{\bf S}}}^{-1}.\label{array:inverse}
\ee
The operator is also anti-Hermitian.


  The array (\ref{array:Sdot}) becomes explicitly
\bea
\dot{{\bf S}}=\gamma\left(\begin{array}{clcr}1,&0,&-iv_z,&-V_+ \\0,&1,&V_-,&iv_z \\-iv_z,&-V_+,&-1,&0 \\V_-, &iv_z,&0,&-1\end{array}\right) \label{array:metricdS}
\eea
This form can be used easily to verify equation (\ref{array:inverse}). The velocities are measured in units of $c$. The quantities $V_\pm=v_y\pm v_x$ .

The fact that $ds^2$ is Lorentz invariant, together with the properties (\ref{array:inverse}) or (\ref{array:dS2}), show that the form of $\dot{{\bf S}}$ is Lorentz invariant. We note that the properties of the transformation matrix do not change under a change in sign of the velocity, so that 
\be
~~~~~~~~~~~~~~~~~~~~\dot{{\bf S}}(-v)=({\dot{{\bf S}}}(-v))^{-1}.\label{array:inverse-}
\ee
However, it is not the case that the matrix with $-{\bf v}$ is the inverse of the matrix with $+{\bf v}$. This has consequences for the transformation properties discussed below. 


Under two cyclic permutations
 $\dot{{\bf S}}$ becomes 
\bea
\dot{{\bf S}}=\gamma\left(\begin{array}{clcr}1&0&-iv_y&-V_+\\0&1&V_-&iv_y\\-iv_y&-V_+&-1&0\\V_-&iv_y&0&-1\end{array}\right)\label{array:PermdS}
\eea
where now 
\be
V_{\pm}=v_x\pm iv_z.\label{eq:Vforms}
\ee
In these expressions $\gamma$ is the Lorentz factor (which should not be confused with the bold face $\bm{\gamma}_a$ matrices )as 
\be
\gamma=\frac{1}{\sqrt{1-{\bf v}^2}}.
\ee

The latter cyclically transformed transformation matrix is useful when extracting from (\ref{eq:trans2}) the Lorentz transformations in two spatial dimensions.  This is because both $v_y$ and $v_z$ are correctly positioned in this matrix to be real velocity components. We shall see this in practice below.

\section{Spinor and  Matrix Transformations} 
\label{sect:LorentzSpinor}
We will proceed mainly with the tranformation (\ref{eq:trans2}) for various four vectors {\bf A}. However the transformation matrix can also operate on a column bi-spinor $\ket$ which we will normally take in the form
\bea
\left(\begin{array}{cl}&\Psi_u\\&\Psi_d\end{array}\right)=\left(\begin{array}{cl}&\psi_0~~\\&\psi_1~~\\&\psi_2~~\\&\psi_3~~\end{array}\right)\equiv\ket.\label{eq:bi-spinors}
\eea 
The quantities $\Psi_u$ and $\Psi_d$ are thus defined as two component spinors. 

\subsection{Spinor Transformations}

 We  have found  privately, proceeding as above, that taking the square root of the Euclidian metric leads naturally to a $2\times 2$ matrix $d{\bf S}$ that acts on two spinors. This matrix has the form (not the $d{\bf S}$ of our current discussion)
\bea
d{\bf S}=\left(\begin{array}{clcr}&dz&dx-idy\\&dx+idy&-dz\end{array}\right).\label{array:dSspace}
\eea
This represents intervals in Euclidian space because  
\be
d{\bf S}^2=ds^2{\bf 1},
\ee
where $ds^2=dx^2+dy^2+dz^2$. It is in fact a rank two spinor representation of a three vector \cite{BLP1971}. The velocity matrix ${\bf \dot S}$ acted as an unconventional ($\det{\bf \dot S}=-1$) spinor transformation and behaved properly under spatial rotations.

The $4\times 4$ matrix that does the same for space-time has been introduced in section (\ref{sect:extract}) and is not simply related to the $2\times 2$ matrix given in (\ref{array:dSspace}). Nevertheless, a slight modification of equation (\ref{array:dSspace}) namely
\bea
{\bf dX}= \left(\begin{array}{clcr}&dt+dz&dx-idy\\&dx+idy&dt-dz\end{array}\right).
\eea
has its determinant equal to the square of the space-time interval. It is a second rank spinor representation of a four vector \cite{BLP1971}. A Lorentz transformation to a frame moving along the $z$ axis with velocity v ($dx=dy=0$) is readily effected by the spinor transformation
\bea
d{\bf X}'=\left(\begin{array}{clcr}&K_-(v)&0\\&0&K_+(v)\end{array}\right){\bf dX}.
\eea
Here the D\"oppler quantities are
\be
K_\pm=\sqrt{\frac{1\pm v}{1\mp v}}.
\ee
These give
\bea
dt'+dz'&=&K_-(v)(dt+dz),\nonumber\\
dt'-dz'&=&K_+(v)(dt-dz),
\eea
from which the standard forms follow. The representation ${\bf X}$ is also subject to cyclic permutation of the $\{xyz\}$.


\subsection{Lorentz Transformation}
\label{sect:Lorentz}
The transformation matrix defined by the square root of the space-time interval  as in section (\ref{sect:properties}) should act on a bi-spinor, but it can also act on a four vector. In this section we consider the implications of the latter action.

We consider  the operation of $\dot{{\bf S}}$ on a four position vector  having the coordinate form
\bea
\bm{X}=\left(\begin{array}{cl}&t\\&ix\\&iy\\&iz\end{array}\right),\label{array:posvector}
\eea
which defines a fixed `event' in the space of an inertial observer $O$. The use of $i$ allows the magnitude of the four vector to be the sum of the squares of the components.

 We compare this with the position vector (which defines the same fixed `event') drawn from the  origin for the moving observer, say O'. The two observers are assumed to have initially ($t=0$) a common origin and their Cartesian axes are  aligned. For completeness we will assume the relative motion  to be in a general spatial direction $\{x,y,z\}$. We use the permuted form (\ref{array:PermdS}) of the transformation matrix to act on ${\bf X}$, although we shall see that this form is more readily interpreted for motion only in the $\{y,z\}$ plane. The latter restriction is not  physically limiting, because one can always arrange the coordinate axes to fit \footnote{Indeed we can, if we wish to assume standard configuration, choose one dimensional motion along, say, the $y$ axis}.  
  
However we must recall the definition of our `gamma' matrices in equations (\ref{array:2ndgammas}). The presence of $i$ in the spatial matrices implies that during the operation on a position vector with the velocity matrix we must use $-i$ times the spatial components in order to have the correct sign of the $\{x,y,z\}$ components of the position vector.  This means that we operate on  ${\bf X(-)}=\{t,-i{\bf X}\}$  with $\dot{{\bf S}}$ (where the transpose is written only for brevity) in order to calculate
\bea
\bm{X'}=\left(\begin{array}{cl}&~~t'\\&ix'\\&iy'\\&iz'\end{array}\right),\label{array:posvector}
\eea

This affects the transformation calculus in that the inverse result (changing as usual the sign of the velocity) requires $\dot{{\bf S}}(-v)$ to operate on 
\bea
\bm{X'(-)}=\left(\begin{array}{cl}&~~t'\\&-ix'\\&-iy'\\&-iz'\end{array}\right),\label{array:posvector}
\eea
in order to give
\bea
\bm{X}=\left(\begin{array}{cl}&t\\&ix\\&iy\\&iz\end{array}\right),\label{array:posvector}
\eea
  Formally, the appropriate inverse matrix applies to either direction of the transformation.


 We carry out the operation $\dot{{\bf S}}{\bf X(-)}$, which should be equal to ${\bf X}'$. This assumes that what we have called the transformation matrix actually transforms vectors between relatively moving observers. We find 
 \bea
\dot{{\bf S}}{\bf X(-)}=\gamma\left(\begin{array}{cl}&t-yv_y+izV_+\\&-ix-iyV_-+zv_y\\&-iv_yt+ixV_++iy\\&tV_-+xv_y+iz\end{array}\right)=\left(\begin{array}{cl}&t'\\&ix'\\&iy'\\&iz'\end{array}\right).\label{array:transposvector}
\eea
It is readily shown that in accordance with the self inverse (e.g. equation(\ref{array:inverse}) it is also true that $\dot{{\bf S}}{\bf X}'={\bf X(-)}$.  

 Remembering the definition of the $V$ quantities ($V_\pm=v_x\pm iv_z$), and confining the motion to the $\{y,z\}$ plane (i.e. $v_x=0$, $x=0$) we find by equating the $\{0,2,3\}$ vector components that the Lorentz transformations take the familiar form
 \bea
 t'&=&\gamma(t+\tilde{\bm{\varpi}}\cdot{\bf v})\nn\\
 \bm{\varpi}'&=&\gamma(\bm{\varpi}-{\bf v}t),\label{eq:Lorentz}
\eea
where 
\be
~~~~~~~~~~~~~~~~~~~~~~~~~~~~~~~~~~~~~~~~~~~~~~\gamma=\frac{1}{\sqrt{1-v_y^2-v_z^2}},
\ee
and 
\bea
\bm{\varpi}=\left(\begin{array}{cl}&0\\&0\\&iy\\&iz\end{array}\right),~~~~~{\bf v}=\left(\begin{array}{cl}&0\\&0\\&iv_y\\&iv_z\end{array}\right).
\eea

The tilde on $\bm{\varpi}$ indicates the transpose.

Note that, although we have set $v_x=0$ and $x=0$, we seem to find  from the $(1)$ component  of the transformation that  the $(1)$ component for $O'$ becomes
\be
~~~~~~~~~~~~~~~~~~~~x'=i\gamma(yv_z-zv_y)\equiv i\gamma \ell_x.\label{eq:angmoment}
\ee
However this term is zero because $y=v_yt$ and $z=v_zt$.

We note that the inverse transformation in differential form 
\be
~~~~~~~~~~~~~~~~~~~~~ d{\bf X}(-)=\dot{{\bf S}}d{\bf X}' \label{eq:diffinverse}
 \ee
 can be considered as a kinematic `solution' for a particle in uniform motion if $d\bm{\varpi}'=0$. In fact  one infers from the  inverse of equation (\ref{eq:Lorentz}) that $d\bm{\varpi}={\bf v}dt$, with $dt=\gamma dt'$.  
 We will see below that the quantum operator version of this equation becomes the Dirac equation for a free particle. In that case electromagnetic forces are included by adding the appropriate canonical momentum to the momentum operator (or appropriate matrices for curved space-time). However, classical particles acted on by (four vector) forces satisfy quite different equations of motion.

We conclude from this section that the transformation matrix operating on the position vector yields the Lorentz transformed position vector. Care must be taken to operate on the same vectors in making this transformation.

\subsection{Four velocity transformation}

We apply the transformation operator to the velocity four vector. Let the velocity of an object for $O$ be
\bea
{\bf u}(-)=\gamma\left(\begin{array}{cl}&~~1\\&~~0\\&-iv_y\\&-iv_z\end{array}\right)\label{array:4u}
\eea

and the velocity for $O'$ is the same form with primes on all quantities and no minus sign. The minus sign in the vector for $O$ is to allow for the $-i$ in the gamma operators, just as previously for the position vector. We suppress any motion along the $x$ axis, both in the object velocity and in the relative velocity of the observers. The velocity components that measure the relative velocity of $O$ and $O'$ appear in the operator $\dot{{\bf S}}$ as $\{V_y,V_z\}$, so the relative motion is also only in the $\{y,z\}$ plane. The gamma factor for the relative motion is denoted $\gamma_V$.

Letting the operator in the form (\ref{array:PermdS}) act on ${\bf u}(-)$ and assuming this yields ${\bf u}'$ gives
\bea
\gamma_V\gamma\left(\begin{array}{cl}&1-v_yV_y-v_zV_z\\&v_zV_y-v_yV_z\\&iv_y-iV_y\\&iv_z-iV_z\end{array}\right)=\gamma'\left(\begin{array}{cl}&~1\\&iv_x'\\&iv_y'\\&iv_z'\end{array}\right)
\eea
The zeroth component yields the transformation of the gamma factor in standard form as 
\be
\gamma'=\gamma\gamma_V(1-{\bf v}\cdot{\bf V}).\label{eq:Lfactor}
\ee
Consequently the $2$ and $3$ components give correctly together (using normal three vector notation)
\be
{\bf v}'=\frac{{\bf v}-{\bf V}}{1-{\bf v}\cdot{\bf V}}.\label{eq:parallelvel}
\ee
The transverse velocity follows as usual by accepting that the transverse  four velocity component $\gamma v_\perp$ is invariant and hence
\be
v_\perp'=\frac{\gamma}{\gamma'}v_\perp=\frac{v_\perp}{\gamma_V(1-{\bf v}\cdot{\bf V})}.\label{eq:vperp}
\ee
However in this case $v_\perp=v_x=0$ so that $v_\perp'$ is also zero.

The term for $O'$ that follows from the $1$ component is 
 \be
v'_x=i\frac{v_yV_z-v_zV_y}{1-{\bf v}\cdot{\bf V}}, \label{eq:vxprime}
\ee
and is  wholly imaginary. Because only the real part is significant for four vectors, this  value may be regarded as zero. This is as expected. In the event that the motion is wholly one dimensional  so that ${\bf v}\parallel {\bf V}$, then this term vanishes in any case.


 
 The inverse transformation of the four velocity is 
 \be
~~~~~~~~~~~~~~~~~~~~ {\bf u}(-)=\dot{{\bf S}}{\bf u}'.\label{eq:velinv}
 \ee
 For a fixed point in the system $O'$, which is moving with the instantaneous four velocity ${\bf u}$  in the $\{y,z\}$ plane, we have the moving system four velocity as 
 \bea
{\bf u}'=\left(\begin{array}{cl}&~~t'\\&~~0\\&~~0\\&~~0\end{array}\right).\label{array:4u'}
\eea
 The $O$ system four velocity is given in equation (\ref{array:4u}), which satisfies equation (\ref{eq:velinv}). In section (\ref{sect:Lorentz}) we remarked that the position vector transformation could be interpreted as the equation of motion for a particle in uniform motion. An additional question is whether an equation of motion can be inferred in this way when the particle is accelerated. In the next section we answer this question in a way that parallels our subsequent derivation of the Dirac equation.
 
 \subsection{Four acceleration transformation}
 
 We can expect, based on previous subsections, that the acceleration vector for $O$ should be related to the acceleration vector for $O'$  by 
 \be
 ~~~~~~~~~~~~~~~~~~~\frac{d{\bf u(-)}}{ds}=\dot{{\bf S}}\frac{d{\bf u'}}{ds},\label{eq:Ainv}
 \ee
 where ${\bf u}$ is the four velocity. The direct transformation follows by exchanging the primed and unprimed  four velocities. 
 
 We expect this expression to yield an equation of motion of an accelerated  $O'$ observer/particle for $O$, provided that we accept the Lorentz transformation between the instantaneous velocities of relatively accelerated `observers'. This reduces the problem  to knowing the accelerations in the frame of $O'$, that is in the proper frame of the accelerated observer/particle. The total derivative with respect to the invariant interval is deliberate, in order to allow for the spatial part of the four velocity to be described in curvilinear coordinates. In Cartesian spatial coordinates the derivative is effectively partial, because the base vectors are already held constant. 
 
 In what follows we assume motion in 2-space ($y,z$), and use the appropriate transformation operator in the form (\ref{array:PermdS}).  In this section we use $u_y$ and $u_z$ as the components of the four velocity. To be explicit,
 \bea
{\bf u}(-)=\left(\begin{array}{cl}&~~\gamma\\&~~0\\&-iu_y\\&-iu_z\end{array}\right),\label{array:43u}
 \eea
 while ${\bf u}'$ is given formally by
 
 \bea
{\bf u}'=\left(\begin{array}{cl}&~~\gamma'\\&~~0\\&~iu'_y\\&~iu'_z\end{array}\right).\label{array:43u'}
 \eea
 The operator (\ref{array:PermdS}) takes the convenient form 
 \bea
\dot{{\bf S}}=\left(\begin{array}{clcr}\gamma&0&-iu_y&-iu_z\\0&\gamma&-iu_z&iu_y\\-iu_y&-iu_z&-\gamma&0\\-iu_z&iu_y&0&-\gamma\end{array}\right)\label{array:PermudS}
\eea
 
 
 We proceed to calculate the prescription (\ref{eq:Ainv}).
 However when calculating $d{\bf u}'/ds$, we take ${\bf u}'={\bf 0}$, $\gamma'=1$ (so that $d\gamma'/ds=0$), but $d{\bf u}'/ds\ne 0$.  That is, ${\bf v}'=0$ but $d{\bf v}'/ds\ne 0$. This corresponds to a transformation to the instantaneous co-moving velocity of the  accelerated particle $O'$. In the appropriate reference frame we recall, either $ds=dt/\gamma$ or $ds=dt'$.
 We obtain the equations of motion in the form
 \bea
 \dot{\gamma}&=&u_y a'_y+u_za'_z,\nonumber\\
 0&=& i(a'_yu_z-a'_zu_y),\nonumber\\
 \dot{u}_y&=&\gamma a'_y,\nonumber\\
 \dot{u}_z&=& \gamma a'_z.\label{eq:motionyz}
 \eea
 
 Here 
 \be
~~~~~~~~~~~~~~~~~~~~~~~~~~~~~~~~~~~~~~~~~~~~~~~a'\equiv\frac{du'}{dt'}=\frac{dv'}{dt'},
 \ee
 because $\gamma'=1$ and recall that the dot in equation (\ref{eq:motionyz}) represents $d/ds=\gamma d/dt$. These equations require only the relativistic force equation for $a'$ to yield the equations of motion of the $O'$ particle for $O$.
 
 The second of these equations comes from requiring $u_x=0$ in addition to $u'_x=0=a'_x$. The condition indicates that ${\bf a}'\parallel {\bf u}$, for consistency. We know that the direction of accelerations in two inertial frames are not necessarily parallel (e.g \cite{PR2011},p101). 
 
 An obvious example sets ${\bf a}'={\bf 0}$, from which we deduce using these equations that ${\bf v}=constant$ and $\gamma=constant$. Thus $v$ is arbitrary up to and including $1$,
 
 A more interesting example is that of hyperbolic motion. We suppose parallel motion and acceleration along the $z$ axis and set $a'_z=a'$. The first and fourth equations of motion become
 \be
 \gamma\dot{\gamma}=u_z\dot{u_z},
 \ee
which is an identity because $\gamma^2=1+u_z^2$.  The second  and third equations are obviously identities and the fourth equation alone becomes
\be
\frac{d(\gamma v_z)}{dt}=a',\label{eq:hyperbolic}
\ee
 where $v_z=dz/dt$. This last equation integrates to give hyperbolic motion as discussed for example in (\cite{PR2011}, pages 102-103).
 
 To include real forces we would normally write the  relativistic equations for ${\bf a}'$. However, this concept does not really exist in the quantum mechanical analogy, nor indeed in gravity. Gravity has been discussed briefly earlier and the Dirac equation will follow. In the case of one dimensional motion, one can use  for a charge $e$ of mass $m_e$ in a parallel electric field the expression for $a'_z$ as 
 \be
 a_z'=\frac{eE'}{m_e}=\frac{eE}{m_e},
 \ee
 which will not be constant in general. A constant field reduces to hyperbolic motion.

\section{The Dirac Equation for Fermions}
\label{sect:Dirac}

In this section we will restore constants and require consistent measurable dimensions. We use the original metric operator (\ref{array:dS}) which has the Dimension of length, and so $\dot{{\bf S}}$ is dimensionless. 
Our objective is to express the Dirac equation in the spirit of the previous sections. That is, to show that it represents the wave function of a free bi-spinor by (quantized) velocity transformation to the rest frame of the particle. To this end, we will abandon the explicit use of the `gamma' matrices in favour of the combined transformation matrix. This gives an unconventional perspective that affords some fresh insight. The equation of motion is more conventionally interpreted in terms of the Hamiltonian operator.
  
We recall that the essential feature of extracting the square root of $ds^2$ with the correct signature  requires a $d{\bf S}$ that is independent of cyclic permutations. Hence, although we find a standard form of the Dirac equations, that form is of course not unique.

We now assume the four component (or two bi-spinors) representation of a free particle (e.g. spin $1/2$, i.e. a fermion) as a complex vector in four dimensional Minkowski space-time\footnote{The vector is actually in Hilbert space but the operator contains the spacetime metric.} We denote the four component complex vector  by $\ket$ so that 
\bea
\ket=\left(\begin{array}{cl}&\psi_0~~\\&\psi_1~~\\&\psi_2~~\\&\psi_3~~\end{array}\right).\label{eq:ket}
\eea 
Because the $\psi_a$ will be complex, the explicit factor $i$ is redundant. Ultimately, we must identify the negative energy solution (and negative momentum) with the antiparticle.

What then could the evolution equation of this four quantity be? We proceed initially by ignoring the quantum operators, so that the argument is basically `classical'. We have seen that the velocity operator gives the evolution of a uniformly moving vector. This may be stated as 
\be
\ket'=\dot{{\bf S}}\ket.\label{eq:kettrans}
\ee
However  a uniformly moving free particle appears as a plane wave. The  $\ket'$, which is translating for $O$, should then be proportional to $\partial_s\ket$ for $O$. This is dimensionally incorrect in the absence of an appropriate constant of proportionality. The  dimensional quantities that we  associate with a free particle experimentally, are the energy and momentum $E$, $p$, the mass $m$, and the fundamental constants $\hbar$ and $c$. However, recalling $E^2/c^2=m^2c^2+p^2$, only three of these are independent in the sense that together they construct a length scale, a time scale, and a mass scale.  The essentially unique choice  is $\{\hbar, m,c\}$. We can then suppose that (the factor $i$ is explained below)
\be
~~~~~~~~~~~~~~~~~~~~~~~~~~~~~~~~~~~~~i\frac{\hbar}{mc}\partial_s\ket=\dot{{\bf S}}\ket,\label{eq:classicalDirac}
\ee
represents an equation of motion for the complex vector. However, by operating on left of this equation with $\dot{{\bf S}}$, using the property (\ref{array:inverse}), and substituting for $\dot{{\bf S}}\ket$ again on the left  ($\dot{\bf S}$ does not depend on $s$ for a uniformly moving particle) gives
\be
~~~~~~~~~~~~~~~~~~~~~~~~~~~~~~~~~~~~~\partial_s^2\ket=-\frac{m^2c^2}{\hbar^2}\ket.
\ee
This shows that our equation can only describe a plane wave if the factor $i$ is added on the left of equation (\ref{eq:classicalDirac}) as we have done. Otherwise the time dependence would be exponential rather than `wave-like'. 
We are therefore forced to use the quantum operator for energy ($i\partial_s$) even in this non relativistic argument, just as is necessary in the Schr\"odinger equation. Here $mc^2\dot{\bf S}$ plays the r\^ole of a Hamiltonian ($ds=cdt$) operator in equation (\ref{eq:classicalDirac}), but the origin of this operator is geometric.

At this stage, operating  on the left of equation (\ref{eq:classicalDirac}) with $\dot{{\bf S}}$, recalling the property (\ref{array:inverse}),  and using the transformation (\ref{eq:kettrans}) in the forward (on the LHS) and backward (on the RHS) senses; shows that the classical equation is strictly Lorentz invariant, because we obtain 
\be
~~~~~~~~~~~~~~~~~~~~~~~~~~~~~~~~~~~~~i\frac{\hbar}{mc}\partial_s\ket'=\dot{{\bf S}}\ket'.
\ee

However we are interested in constructing the Dirac equation, for which we must introduce quantum operators to equation (\ref{eq:classicalDirac}).  We choose to use the coordinate time of each inertial observer  rather than the proper time. This does not change the form of the equation because $ds=cdt/\gamma$ appears on both sides. We  also introduce momentum operators to the transformation matrix as it is given classically in the  form (\ref{array:dS}). Equation (\ref{eq:classicalDirac}) then becomes formally the Dirac equation ($d{\bf S}_q/dt$ is the quantized version of (\ref{array:dS})
\bea
i\frac{\hbar}{mc}\partial_t\ket=\frac{d{\bf S}_q}{dt}\ket.\label{eq:Dirac1}
\eea


We must recall that in equation (\ref{array:dS}) we have used spatial  Dirac/Cifford matrices with an explicit factor $i$. Moreover the signature we have chosen makes this factor $-i$ for the spatial derivatives. Hence the quantized form of (\ref{array:dS}) may be found by : (i) restoring dimensions; (ii) dividing by $dt$; (iii) replacing the classical velocities by classical momenta; and (iv), replacing the classical momentum ${\bf p}$ by the quantum operator $\hbar{\bf\nabla}$. Equation (\ref{eq:Dirac1}) becomes 
\bea
i\frac{\hbar}{mc}\partial_t\ket=\frac{\hbar}{m}\left(\begin{array}{clcr}\frac{mc}{\hbar}&0&-i\partial_z&-\partial_+\\0&\frac{mc}{\hbar}&\partial_-&i\partial_z\\-i\partial_z&-\partial_+&-\frac{mc}{\hbar}&0\\\partial_-&i\partial_z&0&-\frac{mc}{\hbar}\end{array}\right) \ket\label{array:Dirac2}
\eea
The $4\times 4$ matrix, including the factor $\hbar/m$, is the quantized matrix $d{\bf S}_q/dt$, and we have adopted the operator notation
 \be
 \partial_\pm\equiv \partial_y\pm i\partial_x .\label{eq:partial+-}
 \ee
 The matrix $d{\bf S}_q/dt$ in equation (\ref{array:Dirac2}) is skew Hermitian. We also  note that 
 \be
 (\frac{d{\bf S}_q}{dt})^2=c^2(1-\frac{\hbar^2}{m^2c^2}\nabla^2){\bf I}.\label{eq:quantumoperator}
 \ee
  
  Although it can be demonstrated easily from the explicit form (\ref{eq:expDirac}) (see below), we note here that operating on equation (\ref{eq:Dirac1}) with $d{\bf S}_q/dt$ on the left and using  the result (\ref{eq:quantumoperator}) yields the Klein-Gordon equation for $\ket$ as 
  
  \be
  \partial_t^2\ket-c^2\nabla^2\ket=-\frac{m^2c^4}{\hbar^2}\ket.\label{eq:entireKG}
  \ee
We have used equation (\ref{eq:Dirac1}) to eliminate $d{\bf S}_q/dt\ket$ on the left after the matrix multiplication.

  After the introduction of the quantum operators, the Lorentz invariance of the Dirac equation is no longer obvious. However, because this is true for equation (\ref{eq:classicalDirac}), if we recognize that each inertial observer will introduce the same quantum operators then the form of equation (\ref{array:Dirac2}) will be invariant. This will also hold if one introduces Lorentz invariant fields by their gauge or canonical momenta, added to that of the particle.

 Equation (\ref{eq:Dirac1}) differs in detail from the `classical' equation of motion (\ref{eq:Ainv}). The general principle of relativistically transforming  to the proper frame of the wave/particle remains, however. The classical four velocity requires a force description as a function of position and time in order to evolve in spacetime. In the quantum case all of the accessible information regarding a free particle is found in the wave function, together with its evolution in time and space. Spatial dependence enters when the canonical momentum associated with a given field is added to the free momentum operator to give `minimal coupling'. Retaining gauge invariance is the principal purpose of minimal coupling.

 Explicitly, equation (\ref{array:Dirac2}) yields  equations for the four components  of the Dirac wave function  in the form
 \bea
 d_+\psi_0&=&-c\partial_z\psi_2+ic\partial_+\psi_3,\nn \\
 d_+\psi_1&=&-ic\partial_-\psi_2+c\partial_z\psi_3,\nn \\
 d_-\psi_2&=&-c\partial_z\psi_0+ic\partial_+\psi_1,\nn \\
 d_-\psi_3&=&-ic\partial_-\psi_0+c\partial_z\psi_1.\label{eq:expDirac}
 \eea
The new operator is defined as 
\be
d_\pm\equiv \partial_t\pm i\frac{mc^2}{\hbar}.\label{eq:d+-}
\ee

There are many forms of these equations, but this particular version can be found in problem $2$, $ p ~67$ of \cite{BLP1971} after realizing that their matrix notation $\bm{\beta}$ and $\bm{\alpha}$ corresponds to our $\bm{\gamma}_0$ and $-i\bm{\gamma}$. It is what is frequently referred to as the standard form of the Dirac equation.

It is easy to show (by operating either with $d_-$ or $d_+$ and making the appropriate substitutions)  that each component of equations (\ref{eq:expDirac})  satisfy the Klein-Gordon equation in the form ($a=0..3$). 
\be
(\partial_t^2-c^2\bm{\nabla}^2)\Psi_a=-\frac{m^2c^4}{\hbar^2}\psi_a.\label{eq:KG}
\ee
The Klein-Gordon equation only describes the plane wave in spacetime. It does not describe the internal spin degree of freedom that follows from the solution of equation (\ref{eq:Dirac1}). 

 In fact it is more conventional to write equations (\ref{eq:expDirac}) in bi-spinor form, which reveals ultimately the internal spin structure. These become
 \bea
(\frac{i\hbar}{mc^2} \partial_t-1)\left(\begin{array}{cl}&\psi_0\\&\psi_1\end{array}\right)&=&{\bf H}\left(\begin{array}{cl}&\psi_2\\&\psi_3\end{array}\right),\nn \\
(\frac{i\hbar}{mc^2} \partial_t+1)\left(\begin{array}{cl}&\psi_2\\&\psi_3\end{array}\right)&=& {\bf H}\left(\begin{array}{cl}&\psi_0\\&\psi_1\end{array}\right).\label{array:spinDirac}
 \eea

the matrix ${\bf H}$ has the form 
\bea
{\bf H}=-\frac{i\hbar}{mc}\left(\begin{array}{cl}\partial_z,&-i\partial_+\\i\partial_-,&-\partial_z\end{array}\right).\label{array:H}
\eea
One often uses the notation
\bea
\psi_{up}&=&\left(\begin{array}{cl}&\psi_o\\&\psi_1\end{array}\right)\nonumber\\
\psi_{d}~&=&\left(\begin{array}{cl}&\psi_2\\&\psi_3\end{array}\right).
\eea



Expressions (\ref{array:spinDirac}) can be written more simply using the spinors $\Psi_{up}$ and $\Psi_d$ as 
\bea
\frac{i\hbar}{mc^2}d_+\Psi_{up}&=&{\bf H}\Psi_d,\nonumber\\
\frac{i\hbar}{mc^2}d_-\Psi_d&=&{\bf H}\Psi_{up},\label{array:Hamilton}
\eea
from which, operating with $d_-$ in the first equation and $d_+$ in the second, it follows that $\Psi_{up}$ and $\Psi_d$ satisfy the Klein-Gordon equation in the form.
  \bea
(\partial_t^2+\frac{m^2c^4}{\hbar^2})\left(\begin{array}{cl}&\Psi_{up}\\&\Psi_d\end{array}\right)=c^2\nabla^2\left(\begin{array}{cl}&\Psi_{up}\\&\Psi_d\end{array}\right),
\eea

One relativistic spinor feeds off the other, which is related to the zitterbewegung (cf \cite{Pen2005}, chpt. 25) of a spin $1/2$ particle.  Moreover the free electron (or fermion) velocity is $c$, which leaves only a mean velocity to be sub luminal.  The $\pm 1$ that occurs on the left of equations (\ref{array:spinDirac}) allows for the rest mass energy that is not present in the Euclidian case. 

 The Klein-Gordon equation gives the spacetime dependence of the spinor, although it contains more possibilities than the first order Dirac equation which adds constraints. The spinor behaviour is introduced by considering a particle in its rest frame and solving for two component factors that describe spin up and spin down. 
We note that for a spinor at rest (so that ${\bf H}$  and hence $\nabla^2$ vanishes), there is a wave function oscillation with frequency 
\be
\nu=\frac{mc^2}{\hbar}.
\ee

 A translating free particle has the solution (for each component)
\be
\psi_A=s_A\exp{(-i(\frac{Et}{\hbar}-\frac{{\bf p}\cdot{\bf x}}{\hbar}})),\label{eq:free particle}
\ee
where $s_A$ denotes a pure spinor factor. The expected dispersion relation $E^2-c^2{\bf p}^2=m^2c^4$ follows, and the energy can be either positive or negative. The negative energy leads to one spinor describing the positron (see e.g. \cite{AH1982}).



\subsection{Electromagnetic Field}
\label{sect:magneticfield}

To introduce the electromagnetic force one uses  gauge invariant operators in place of the temporal and spatial derivatives, according to 
\be
\partial^\mu\rightarrow \partial^\mu+\frac{ieA^\mu}{\hbar c},\label{eq:gaugeoperators}
\ee
where $e$ is the algebraic particle charge and $A^\mu$ is the electromagnetic four potential $(\Phi,{\bf A})$. We use Gaussian units. In equation (\ref{eq:gaugeoperators}) 
\be
\partial^\mu=(\frac{1}{c}\partial_t, -{\bf \nabla}).
\ee 
and so 
\bea
\partial_t&\rightarrow& \partial_t+\frac{ie}{\hbar}\Phi,\nonumber\\
-{\bf\nabla}&\rightarrow&-{\bf\nabla}+\frac{ie}{\hbar c}{\bf A}.\label{eq:expEMtrans}
\eea
The free particle momentum operator, $p^\mu=i\hbar\partial^\mu$, becomes 
\be
p^\mu=i\hbar(\partial^\mu+\frac{ie}{\hbar c}A^\mu)
\ee
or explicitly
\bea
p^0&=& \frac{i\hbar}{c}\partial_t-\frac{e}{c}\Phi,\nonumber\\
{\bf p}&=& -i\hbar{\bf \nabla}-\frac{e}{c}{\bf A}.
\eea

For example, if we quantize the expression (\ref{array:PermdS}) to obtain $d{\bf S}_q/dt$ the matrix ${\bf K}\equiv (m/\hbar)d{\bf S}_q/dt$ becomes (to obtain the matrix in the permutation of equation (\ref{array:Dirac2}) let $x\rightarrow y\rightarrow z\rightarrow x$) 

\bea
{\bf K}=\left(\begin{array}{clcc}-\frac{imc^2}{\hbar}-\frac{ie}{\hbar}\Phi &~~~~~~~0&-c(\partial_y-\frac{ie}{\hbar c}A_y)&ic(\partial_+ -\frac{ie}{\hbar c}A_+)\\0&-\frac{imc^2}{\hbar}-\frac{ie}{\hbar}\Phi&-ic(\partial_--\frac{ie}{\hbar c}A_-)&c(\partial_y-\frac{ie}{\hbar c}A_y)\\ -c(\partial_y-\frac{ie}{\hbar c}A_y)&ic(\partial_+ -\frac{ie}{\hbar c}A_+)&\frac{imc^2}{\hbar}-\frac{ie}{\hbar}\Phi&0\\ -ic(\partial_--\frac{ie}{\hbar c}A_-)&c(\partial_y-\frac{ie}{\hbar c}A_y)&0&\frac{imc^2}{\hbar}-\frac{ie}{\hbar}\Phi\end{array}\right)\label{array:KEM}
\eea
Here 
\be
A_\pm=A_x\pm iA_z,~~~~~~~~~~~~~~\partial_\pm=\partial_x\pm i\partial_z.
\ee

To pursue a solution we may use the pair of spinor equations (\ref{array:Hamilton}).  When extended to include the electromagnetic field, the operators in equations (\ref{array:Hamilton}) become 
\bea
&\frac{i\hbar}{mc^2}d_{\pm}+\frac{ie}{\hbar}\Phi\leftarrow \frac{i\hbar}{mc^2}d_{\pm},\nonumber\\
-&\frac{i\hbar}{mc}\left(\begin{array}{clcr}\partial_z-\frac{ie}{\hbar c}A_z~~,~~ -i~(\partial_+-\frac{ie}{\hbar c}A_+)\\i(\partial_- -\frac{ie}{\hbar c}A_-)~,~-(\partial_z-\frac{ie}{\hbar c}A_z)\end{array}\right)\leftarrow {\bf H}\label{array:quantumHamilton}
\eea
Now we have returned to the notation 
\bea
\partial_{\pm}&=&\partial_y\pm i\partial_x,\nonumber\\
A_{\pm}&=& A_y\pm A_x.\label{eq:notation}
\eea

As a familiar example we can assume a pure external magnetic field that is time independent acting on the charge. Then we apply the resulting operator $d_-$ (equation(\ref{array:quantumHamilton}) and \ref{eq:d+-}) to the first of equations (\ref{array:Hamilton}) to obtain
\be
\frac{i\hbar}{mc^2}d_-(d_+\Psi_{up})=d_-({\bf H}\Psi_d).
\ee
Applying the operators in sequence as indicated, this equation can be written
\be
\frac{1}{c^2}(\partial_t^2+\frac{m^2c^4}{\hbar^2})\Psi_{up}={\bf H} (d_-\Psi_d)={\bf H}^2\Psi_{up},\label{eq:psiup}
\ee
where we remember that $\Psi_{up}$ has two components. Here  and below ${\bf H}$ is as in equations (\ref{array:quantumHamilton}) but without the factor $i\hbar/mc$. The last equation makes use of the second of equations (\ref{array:Hamilton}). 

 Following a similar procedure starting with the operator $d_+$ applied to the second of equations (\ref{array:Hamilton}), we obtain the same equation for $\Psi_d$.  The operator ${\bf H}^2$ can be written after a straightforward but tedious calculation (in a gauge with zero  divergence of ${\bf A}$)
 \be
 {\bf H}^2=\frac{e}{\hbar c}(\bm{\sigma}\cdot ({\bf B}+{\bf A}\times{\bf \nabla}))+(\nabla^2-\frac{ie}{\hbar c}{\bf A}\cdot{\bf \nabla}-(\frac{e}{\hbar c})^2{\bf A}^2){\bf 1}.\label{eq:H(EM)}
 \ee
 This completes equation (\ref{eq:psiup}) and the same equation applies to $\Psi_d$.  When the magnetic field is zero, equation (\ref{eq:psiup}) reduces to the Klein-Gordon equation.
 
 
 For a positive energy particle nearly at rest in the magnetic field, equation (\ref{eq:psiup}) with the ans\"atz $\Psi_{up}\propto e^{-iE t/\hbar } \psi_{up}$ becomes,  
 \be
 \frac{1}{c^2}(-\frac{E^2}{\hbar^2}+\frac{m^2c^4}{\hbar^2})\psi_{up}=\big[\frac{e}{\hbar c}(\bm{\sigma}\cdot ({\bf B}+{\bf A}\times{\bf \nabla}))+(\nabla^2-\frac{ie}{\hbar c}{\bf A}\cdot{\bf \nabla}-(\frac{e}{\hbar c})^2{\bf A}^2){\bf 1}\big]\psi_{up}
 \ee
 With $E=mc^2+mv^2/2$ this last equation after factoring becomes ( and $mc^2\gg mv^2$)
 \be
 -(m\frac{v^2}{2c^2})(2mc^2)\psi_{up}=\frac{e\hbar}{c}(\bm{\sigma}\cdot {\bf B})\psi_{up}.
 \ee
  Consequently (see e.g. \cite{Peacock}), the electron magnetic moment has the correct value ($e\hbar/2m$).  The discussion is the same for a negative energy particle.

The story continues now into familiar territory for spin $1/2$ particles, but we leave it here.  We have given a presentation of familiar things from a slightly different perspective.  Equation (\ref{eq:Dirac1}) is a quantized transformation to a moving free wave/particle. Forces, such as the Electromagnetic force, enter by using the canonical momentum. The gravitational force may be introduced by adjusting the Dirac/Clifford matrices to give the appropriate interval. This is certainly possible for metrics of high symmetry.

\section{Discussion}
\label{sect:discussion}

In this paper we  used the Dirac/Clifford method of extracting the square root of the metric interval in Minkowski spacetime, which may be extended to a curved spacetime of high symmetry.  

This procedure led us to a matrix  $d{\bf S}$ which  operates on classical four vectors and bi-spinors. 
The transformation version of this matrix  $\dot{\bf S}$, Lorentz transforms correctly the same four vector, as identified by two relatively moving inertial observers. The matrix is form invariant for each inertial observer, which is compatible with it being its own inverse. The matrix $\dot{\bf S}$ can be used to define an `equation of motion' for a free, classical particle. It does this by describing the time dependence as the transformation of the four acceleration to a uniformly moving  particle frame. This  rather peculiar approach  is discussed as a preliminary to the nature of the Dirac equation. That equation follows when the quantized velocity matrix acts on Dirac four spinors.  The transformation matrix (times $mc^2$) is the Hamiltonian/Energy  operator as well as being a Lorentz transformation.

By quantizing the transformation matrix, we derive the Dirac equation for a four  component complex wave function. There is a close correspondence to the classical equation of motion, which in that case is found by transforming the four acceleration from the rest frame of the particle to the observer. Each component of the complex wave function satisfies the Klein-Gordon equation, not explicitly as a requirement, but as a natural consequence of the interval square root procedure. The Dirac equation imposes additional constraints on the solutions of the Klein/Gordon equation. 


The interaction of a quantum particle with a given field, such as an electromagnetic field or a gravitational field, may be included with the introduction of the appropriate momentum operator. This extension is found by including  in the quantized transformation matrix, the canonical momentum in the electromagnetic case; or an appropriate space-time metric in the gravitational case. In the gravitational case the momentum operator must be given in the appropriate spatial geometry.  The application to spherically symmetric, static metrics is readily formulated.

As a familiar example we have derived the Dirac equation for a charged Fermion a time independent magnetic field. It is in a slightly unconventional approach, but the correct form of the electron magnetic moment is seen to appear in a non relativistic, two-scale limit. 

The main accomplishment of this investigation based on the square root of a geometrical metric, is to illustrate the intimate connection between spacetime geometry and the quantized  relativistic equations of motion, especially for a free wave//particle. There is a kind of `metric imperative' acting as an automatic constraint, similar in fact to that provided by the relativistic energy \cite{Harris1972}. 
A related idea has been introduced previously \cite{K2008}, \cite{BK2008}, but with a more abstract emphasis than ours and less distinction between the relativistic and non relativistic metrics.

One can use the Euclidian spatial metric in a similar way  to obtain the Pauli equation and the Minkowski metric led to the Dirac equation in this study. More complicated metrics may lead to unexpected equations of motion and results (e.g. \cite{DDR1976}).

After  beginning  this project, I learned of a work by De-Sheng Li \cite{Li2014}, that has similar inspiration to this work. However a full application to quantum field theory was attempted there. The direct intuitive route that leads directly to the Dirac equation and to the end of single particle theory that I have followed, is still relatively unfamiliar.

  The approach does not yet produce any new physical results. However, there may be simplifications and hints of new insights.  
 Some more understanding of the relativistic origin of the Dirac equation has been gained.  The action and origin of the various operators seems somewhat more transparent than in the conventional representations.
\bigskip

{\bf Competing Interests: The author declares there are no competing interests.}
\bigskip

{\bf Data generated or analyzed during this study are provided in full within the published article.}

\end{document}